\documentclass[seceq]{ptptex}

\usepackage{graphicx}
\usepackage{wrapft}

\usepackage{bm}

\title{
QCD phase diagram and the critical point%
}

\author{
Mikhail \textsc{Stephanov}%
}

\inst{
Department of Physics, University of Illinois, Chicago, IL 
60607-7059\\
and\\
RIKEN-BNL Research Center, Brookhaven National Laboratory,
Upton, NY 11973
}

\abst{
The recent progress in understanding the
QCD phase diagram and the physics of the QCD critical point
is reviewed.
}

\begin{document}

\maketitle

\section{Preamble}

Quantum Chromodynamics is one of the most remarkable theories of Nature. Its
mathematical foundations are concise, yet the
phenomenology which the theory describes is broad and diverse.
QCD phenomenology at finite temperature and baryon number density
is one of the least explored regimes of the theory.
There are several experimental windows into this regime.
One is the physics associated with the interior of neutron stars.
Another, which is the subject of the ongoing and planned experimental
programs, is the physics of heavy ion collisions.

This report is a review of recent developments in our understanding,
mostly theoretical, but also experimental, of the phase diagram of QCD.  There
exist a number of excellent recent reviews
\cite{RajagopalWilczek,Alford,Schafer,Rischke,Hong,Muroya-review}
which discuss many of the questions addressed here as well as the
related material not covered in this report. The report
focuses on the physics of the critical point of QCD and its
search. There are many open questions in this field, and some of the
theoretical as well as experimental results and expectations discussed
here might not
hold under further scrutiny. Nevertheless it is hoped that this report
will provide a useful contemporary guide to both theorists and
experimentalists entering the field as well as a stimulating reading
to the field's experts.

\section{What is the QCD critical point?}

\subsection{The phase diagram}

\begin{figure}[htbp]
  \begin{center}
    \includegraphics[width=\textwidth]{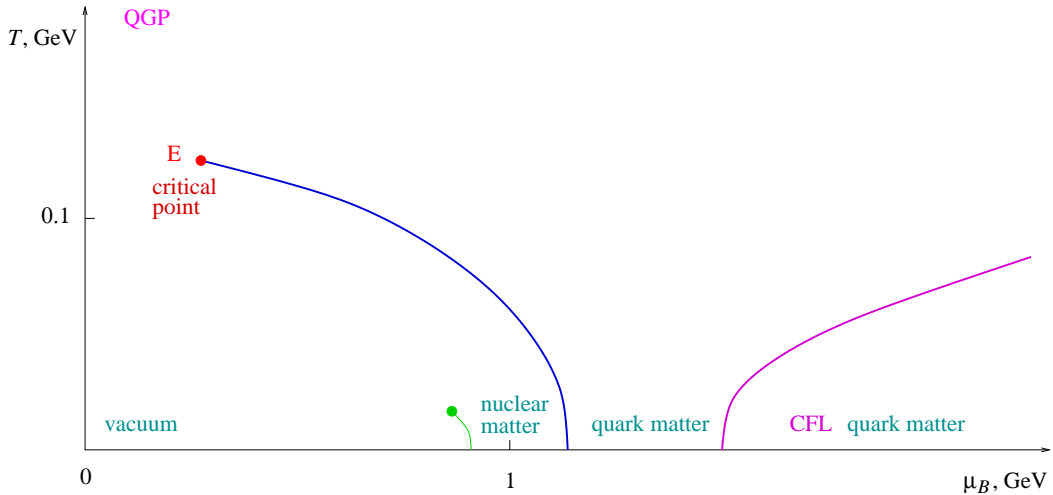}
    \caption{QCD phase diagram}
    \label{fig:pdqcd}
  \end{center}
\end{figure}

Fig. \ref{fig:pdqcd} shows a sketch of the QCD phase diagram
as it is perceived by a modern theorist. By a phase diagram we shall
mean the information about the location of the phase boundaries
(phase transitions) as well as the physics of the phases that these
transitions delineate. The phase transitions are the thermodynamic
singularities of the system. The system under consideration is a
region (in theory, infinite) occupied by strongly interacting matter,
described by QCD, in thermal and chemical equilibrium, characterized
by the given values of temperature $T$ and baryo-chemical potential
$\mu_B$. In practice, it can be a region in the interior of a neutron
star, or inside the hot and dense fireball created by a heavy ion collision.

On the phase diagram,
the regime of small $T$ and large $\mu_B$ is of relevance to neutron
star physics. Because of low temperature, a very rich spectrum of
possibilities of ordering  can be envisaged. The line
separating the Color-Flavor-Locked (CFL) phase, predicted in Ref.~\citen{CFL}, 
from the higher temperature
disordered phase (quark-gluon plasma, or QGP) is the most simplified
representation of the possible phase structure in this region.
This regime is also of particular {\em theoretical} interest because
analytical controllable calculations are possible, due to asymptotic
freedom of QCD. The reader is referred to the reviews 
\cite{RajagopalWilczek,Alford,Schafer,Rischke,Hong}
which cover the recent developments in the study of this domain
of the phase diagram.

The region of the phase diagram more readily probed by the heavy ion
collision experiments is that of rather large $T\sim 100$ MeV,
commensurate with the inherent dynamical scale in QCD, and small to
medium chemical potential $\mu_B\sim0-600$ MeV. Theorists expect that 
this region has an interesting
feature -- the end point of the first order phase transition line,
the critical point marked $E$ on Fig. \ref{fig:pdqcd}. 
The physics of
this point is the focus of the review.

\subsection{Why should there be a critical point?}
\label{sec:argument}

The argument (which is not a proof) that the point $E$ must
exist is short, and is based on a small number of reasonable assumptions.
The two basic facts that it relies on
are as follows:

(1) The temperature driven transition at zero $\mu_B$ is not a
thermodynamic singularity. Rather, it is a rapid, but smooth,
crossover from the regime describable as a gas of hadrons, to
the one dominated by internal degrees of freedom of QCD --
quarks and gluons.  This is the result of finite $T$ lattice
calculations \cite{columbia}.

(2) The $\mu_B$ driven transition at zero $T$ is a first order
phase transition. This conclusion is less robust, since the
first principle lattice calculations are not controllable in this
regime (naive Euclidean formulation of the theory suffers from the
notorious sign problem at any finite $\mu_B$). Nevertheless
a number of different model approaches%
\cite{Asakawa,Barducci:1989,Barducci:1993,BergesRajagopal,Halasz,Scavenius,Antoniou,HattaIkeda} (see Section \ref{sec:theory}) 
indicate that the transition in this region
is strongly first order.

(3) The last step of the argument is a logical product of (1) and (2).
Since the first order line originating at zero $T$ cannot end
at the vertical axis $\mu_B=0$ (by virtue of (1)), the
line must end somewhere in the midst of the phase diagram.

The end point of a first order line is a critical point of the second
order.  This is by far the most common critical phenomenon in
condensed matter physics.  Most liquids possess
such a singularity, including water. The line
which we know as the water boiling transition ends at pressure $p=218$ atm
and $T=374^\circ$C.  Along this line the two coexisting phases (water and
vapor) become less and less distinct as one approaches the end point
(the density of water decreases and of vapor increases),
resulting in a single phase at this point and beyond.

In QCD the two coexisting phases are hadron gas (lower $T$), and
quark-gluon plasma (higher $T$). What distinguishes the two phases? As
in the case of water and vapor, the distinction is only quantitative,
and more obviously so as we approach the critical point.  Rigorously,
there is no good order parameter which could distinguish the two phases
{\em qualitatively}. The chiral condensate,
$\langle\bar\psi\psi\rangle$, which comes closest to being an order
parameter, is non-zero in both phases because of the finite bare quark
mass. Deconfinement, although a useful concept to discuss the
transition from hadron to quark-gluon plasma, strictly speaking, does
not provide a good order parameter. Even in vacuum ($T=0$) the confining
potential cannot rise infinitely -- a quark-antiquark pair inserted
into the color flux tube breaks it. The energy required to separate
two test color charges from each other is finite if there are light
quarks.

\subsection{Critical or tricritical?}

There is an idealization of QCD where the distinction between the
hadron gas and quark gluon plasma is sharp. It describes the world with
massless quarks $m_q=0$. In this limit of QCD with 2
massless quarks (up and down) the chiral symmetry SU(2)$_V\times$SU(2)$_A$
is exact. Although interactions respect this
symmetry, it is spontaneously broken in the QCD vacuum to SU(2)$_V$, and the
Goldstone theorem demands 3 massless bosons -- the pions.  This
breaking is a result of nonperturbative dynamics in QCD (instantons
provide a natural mechanism).

The breaking of the global
symmetry, such as the chiral symmetry, can be thought of as
establishment of the long-range order in the vacuum. It is the order which
dictates the preferred SU(2)$_A$ direction for all points in space, 
and over which the pions are quantized ``ripples''.
At sufficiently high $T$, the order is melted as in any other
such system (compare, e.g., to the disordering of the
ferromagnet at Curie temperature). The chiral symmetry
is restored. The two phases must be separated by a thermodynamic
singularity -- a phase transition. 

This argument can be made more
rigorous by considering the order parameter,
$\langle\bar\psi\psi\rangle$, the expectation value, or the
condensate, of a field transforming non-trivially under the broken
symmetry. Consider a fixed value of $\mu_B$, e.g., $\mu_B=0$.
The chiral condensate $\langle\bar\psi\psi\rangle$ 
as a function of $T$
is identically zero (by
symmetry) for all temperatures above some value $T_c$  and 
is nonzero function of $T$ below (symmetry breaking). 
Such a function cannot be analytic.
A singularity must occur at $T_c$.

\begin{figure}[htbp]
  \begin{center}
    \includegraphics[width=.7\textwidth]{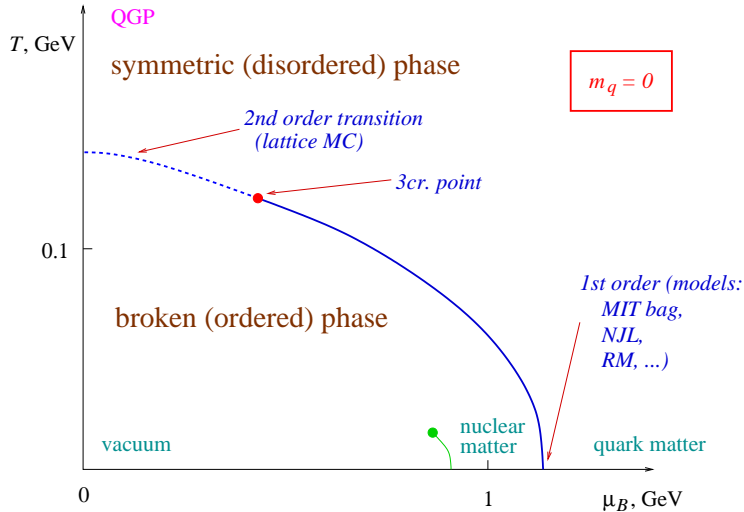}
    \caption[]{Phase diagram of QCD with two {\em massless\/} quarks.
    The chiral symmetry order parameter {\em qualitatively} distinguishes
    two phases: $\langle\bar\psi\psi\rangle\ne0$ in the broken phase
    and $\langle\bar\psi\psi\rangle=0$ in the symmetric phase.}
    \label{fig:pdm=0}
  \end{center}
\end{figure}

In QCD, lattice calculations show that this singularity is a
second order phase transition if $T_c$ is approached at $\mu_B=0$.%
\cite{Gottlieb,Bernard,Fukugita,Mawhinney,Zhu,Iwasaki,Bernard-2,columbia}
At other values of $\mu_B$ the critical temperature $T_c$ is
different, but the line
of transitions $T_c(\mu_B)$ cannot terminate, since any path from
the vacuum $T=\mu_B=0$ to the high $T$ phase must cross a singularity.
Somewhere in the midst of the phase diagram the order of the
transition should change to first order (according to point (2) of
Section \ref{sec:argument}).%
\footnote{We also assume that there is
only one transition between the broken and symmetric phases.}
The point where this happens is the {\em tricritical point}.
The resulting phase diagram is illustrated in Fig. \ref{fig:pdm=0}.

\begin{wrapfigure}{r}{\halftext}
  \begin{center}
    \includegraphics[width=.35\textwidth]{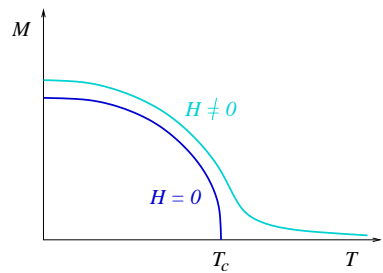}
    \caption{The order parameter vs temperature in a Curie
      ferromagnet with zero and non-zero applied magnetic field. 
      In QCD, the chiral order parameter
      $\langle\bar\psi\psi\rangle$ behaves similarly 
as a function of $T$ at $m_q=0$ and
$m_q\ne0$.}
    \label{fig:ferromag}
  \end{center}
\end{wrapfigure}

Once the quark mass $m_q$ is turned back on, the distinction between
the symmetric and broken phases is blurred, and the second order phase
transition is replaced by a smooth crossover.  The situation is
analogous to the ferromagnet --- an arbitrary small magnetic field
(the analog of $m_q$) smooths away the Curie singularity (Fig.
\ref{fig:ferromag}).  The first order phase transition, on the other
hand, is associated with a finite discontinuity of the order parameter and 
cannot be removed by an arbitrarily small perturbation $m_q\ne0$. 
Thus we arrive back at the diagram in Fig. \ref{fig:pdqcd}.

\begin{figure}[htbp]
  \begin{center}
    \includegraphics[width=.8\textwidth]{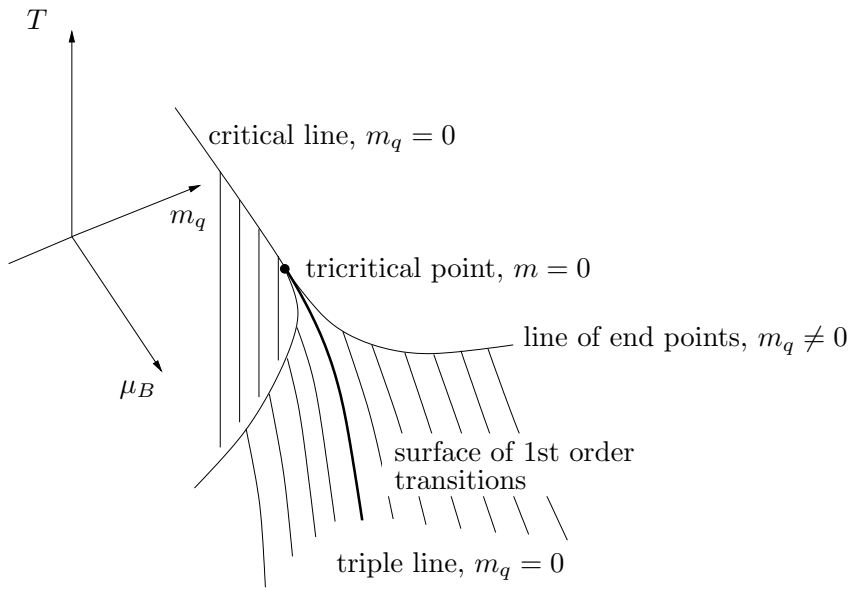}
    \caption{A three-dimensional view ($T$, $\mu_B$, $m_q$) of the QCD 
phase diagram near the tricritical point.}
    \label{fig:wings}
  \end{center}
\end{figure}

It is also useful to take a look at Fig. \ref{fig:wings}, where the
2-dimensional $T\mu_B$ phase diagram is extended to 3-dimensions by
adding the quark mass $m_q$ as the third axis. One can see that the
second order transition line at $m_q=0$ does not extend into
$m_q\ne0$. This line can be seen as a boundary of the coexistence 
surface of the two spontaneously broken phases with
$\langle\bar\psi\psi\rangle$ of opposite signs.  A first order line
ending at a critical point, on the other hand, exists for all nonzero
(small) $m_q$, thus making up a surface which looks like two wings in
Fig.~\ref{fig:wings}. The tricritical point can be seen as the end of
a first order line where 3 phases coexist (line of triple points).

\begin{figure}
  \begin{center}
    \includegraphics[width=.55\textwidth]{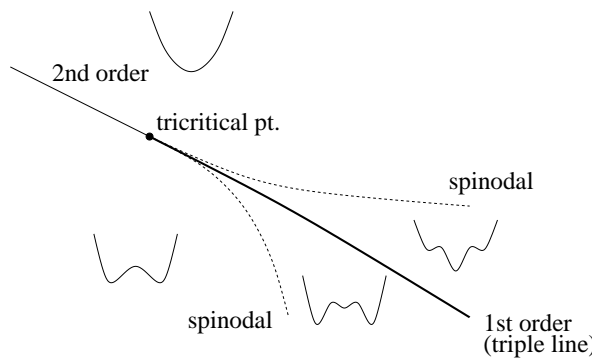}
    \caption{Illustration of the shape of the effective
potential for the chiral order parameter near the tricritical point
in the $m_q=0$ plane. Two additional (spinodal) lines, not present in
Fig.~\ref{fig:wings} indicate the boundary of the existence of
metastable minima.}
    \label{fig:3min}
  \end{center}
\end{figure}

Another useful sketch is made in Fig \ref{fig:3min}. It shows, in a
schematic way, the shape of the effective potential in various
regions around the tricritical point. One can see that the three minima,
which are equally deep on the triple line fuse into one minimum at the
tricritical point.

\subsection{Critical behavior: static and dynamic universality class}

Determining properties of QCD (equation of state, correlation
functions, etc.) near the critical point is difficult, for the same
reason as it is difficult to find the location of the critical point
(see next Section). However, as it is the case for any critical point,
singular properties, such as critical exponents, can be determined
using universality arguments. 

According to the scaling postulate, central to the theory of critical
phenomena,\cite{Amit} all singular contributions to the thermodynamic
quantities are powers of the correlation length $\xi$, which diverges
at the critical point. These powers, or critical exponents, are
universal, in the sense that they depend only on the degrees of
freedom in the theory and their symmetry, but not on the other
details of the interactions. Very different physical systems may belong
to the same universality class, as far as their critical behavior is concerned.

One should distinguish static and dynamic universality classifications
\cite{HH}. From the point of view of static critical phenomena, the
QCD critical point falls into the universality class of the Ising
model. This is a consequence of the fact that at $m_q\ne0$ no symmetry
remains which would require the order parameter to have more than just
one component. The field theory which describes the static critical 
behavior, the one-component $\phi^4$ theory in 3 dimensions,
 has the critical exponents of the Ising model.
\footnote{As another example, consider any of the critical points on the
2nd order line at $m_q=0$ on Fig.~\ref{fig:pdm=0}. Because of the O(4)$\sim$
SU(2)$_V\times$SU(2)$_A$ symmetry, which is restored at this critical
point, the order parameter must carry 4 components --- sigma and 3 pions
$(\sigma,\bm \pi)$. The resulting field theory describes the
universality class of the $O(4)$
ferromagnet.\cite{PisarskiWilczek,RajagopalWilczek-2,Rajagopal}}

What is the nature of this order parameter? It can be taken as the
value of the chiral order parameter $\bar\psi\psi$ (often called
$\sigma$), since it is distinct in two phases coexisting across the
first order phase transition terminating in the critical point. As a
result, the static (equal-time) correlation function
$\langle\bar\psi\psi(\bm x)\bar\psi\psi(\bm y)\rangle$ develops
divergent correlation length:
\begin{equation}
\langle\bar\psi\psi(\bm x)\bar\psi\psi(\bm 0)\rangle_{\rm c}
\sim \left\{
\begin{array}{ll}\displaystyle
  \frac1{|\bm x|^{1+\eta}}, &\qquad |\bm x|\ll\xi;\\
\\
e^{-|\bm x|/\xi}, &\qquad |\bm x|\gg\xi;
\end{array}
\right.
\end{equation}
where $\langle\bar\psi\psi(\bm x)\bar\psi\psi(\bm 0)\rangle_{\rm
  c}\equiv\langle\bar\psi\psi(\bm x)\bar\psi\psi(\bm 0)\rangle-
\langle\bar\psi\psi\rangle^2$.  
The correlation length diverges, $\xi\to\infty$, as we
approach the critical point. For the Ising universality class
 $\eta\approx 0.04$.

Another interesting quantity, both from theoretical and
experimental points of view, is the baryon
number density $n_B(\bm x)$. Because symmetry (or, rather, the
absence of such) allows mixing of $n_B(\bm x)$ with $\bar\psi\psi(\bm x)$,
the divergence of the baryon number susceptibility is related to
the divergence of the correlation length $\xi$:
\begin{equation}\label{Bsusc}
  \frac{\partial n_B}{\partial \mu_B}
=\int d^3{\bm x}
\langle n_B(\bm x) n_B(\bm 0)\rangle_{\rm c}
\sim \int d^3{\bm x}
\langle\bar\psi\psi(\bm x)\bar\psi\psi(\bm 0)\rangle_{\rm c}
\sim \xi^{2-\eta}.
\end{equation}

The baryon number density also jumps across the first order phase
transition. One can equally well use $n_B$ as the degree of freedom in
the effective theory near the critical point, or any linear
combination of $\bar\psi\psi$ and $n_B$ (or any other field which can mix
with $\bar\psi\psi$) which is discontinuous across the first order phase
transition. Regardless of the choice, there is only one order
parameter, as far as the static critical behavior is concerned.

The situation is similar but a little more complicated if one
considers {\em dynamic} critical behavior, e.g., the singularities of
kinetic coefficients, etc.  The scaling postulate is similar in
this case, but the universality classes are now determined by the
degrees of freedom which define the effective {\em hydrodynamic}
theory near the critical point.\cite{HH} In this case the fundamental
difference between $\bar\psi\psi$ and $n_B$ fields is that
the latter is a {\em conserved} density. The hydrodynamic
equations for $n_B$ are diffusive, while the dynamics of $\bar\psi\psi$ is
relaxational. Because the two modes mix, there is, again, only one
independent hydrodynamic variable, and it is diffusive.%
\cite{Fujii,DynScaling} This mode involves fluctuations of both $\bar\psi\psi$
and $n_B$ in a fixed proportion.  The fluctuations of $\bar\psi\psi$ alone
relax on a finite time scale even at the critical point.%
\footnote{A related observation, that the sigma {\em pole} 
mass does not vanish at the critical point in the large-$N$ 
NJL model, was made 
in Ref. \citen{Scavenius} and confirmed in Ref.  \citen{Fujii}.}

The complete hydrodynamic theory near the critical point must also
involve the energy and momentum densities.  Once the hydrodynamic
equations are written down, and the mixing between $\bar\psi\psi$, $n_B$ and
the energy density is taken into account, one finds the theory
equivalent to the one describing the liquid-gas phase transition,
model H in the classification of Ref. \citen{HH}.  One consequence of
this theory, interesting from phenomenological point of view, is the
vanishing of the baryon number diffusion rate at the critical point:
$D\sim\xi^{-x_D}$, with exponent $x_D\approx 1$.\cite{DynScaling}

\section{Where is the critical point? Theory}

Theoretically, finding the coordinates $(T,\mu_B)$ of the
critical point is a well-defined task. We need to calculate
the partition function of QCD and find the singularity corresponding
to the end of the first order transition line. The Lagrangian of QCD
is known, and the partition function is given by a path integral
of the exponent of the QCD action, after Wick rotation to the
Euclidean space (with imaginary time compactified on a torus of
circumference $1/T$). 

Of course, calculating such an infinitely
dimensional integral analytically is beyond our present abilities
(perturbation theory is not an option here, in the relevant
region of $T$ and $\mu_B$). We are thus left with numerical
methods, i.e., lattice Monte Carlo simulation. At zero $\mu_B$
this method allows us to determine the equation of state of QCD
as a function of $T$ and reach the conclusion (1) in Section
\ref{sec:argument}. However, at finite $\mu_B$ the Nature guards its secrets
better.

\subsection{Importance sampling and the sign problem}

The notorious sign problem has been known to lattice experts since
early days of this field. Calculating the partition function using
Monte Carlo method hinges on the fact that the exponent of the
Euclidean action $S_E$ is a positive definite function of its 
variables (values of the fields on the lattice). This allows
one to limit calculation to a relatively small set of field configurations
randomly picked with probability proportional to the value of $\exp(-S_E)$.
The number of such configurations needed to achieve reasonable
accuracy is vastly smaller than the total number of possible
configurations. The latter is exponentially large in the size $V$ 
of the system, 
or, the number of the degrees of freedom: $\exp({\rm const}\cdot V)$.
The method, also known as importance sampling, utilizes the fact
that the vast majority of these configurations contribute a
tiny fraction because of the exponential suppression by $\exp(-S_E)$.
Only configurations with sizable $\exp(-S_E)$ are important.

At $\mu\neq0$ the action $S_E$ is complex. What configurations are
important then? A number of ways to circumvent
the problem have been tried. For example,
using the modulus of $\exp(-S_E)$ as a measure of importance, or
the value of $\exp(-S_E)$ at {\em zero} $\mu_B$. Unfortunately,
none work, at present.

\subsection{The overlap problem}

 For the latter choice, $\exp(-S_E)|_{\mu_B=0}$, the problem can be
understood physically and is known as the {\em overlap problem}. The
important configurations at $\mu_B=0$ are different than those of
${\mu_B\ne0}$. How bad is this quantitatively?  At finite volume, even
at $\mu_B=0$, the configurations important for ${\mu_B\ne0}$ pop up,
but with a small probability.  This probability is exponentially
small as volume $V\to\infty$: $\exp(-{\rm const}\cdot V)$. When we
calculate the partition function using this method, we correct for
this by multiplying the contribution of these rare configurations by
the factor $\exp(+S_E|_{\mu_B=0}-S_E)$. The procedure is termed {\em
reweighting}.\footnote{The reweighting method in application
    to finite $\mu_B$ calculations is known as the
 ``Glasgow method'' (reviewed in Ref. \citen{Barbour-review}).}
 The reweighting factor is exponentially large as
$V\to\infty$ --- both the magnitude and the complex phase are
$\exp({\rm const}\cdot V)$.  Fluctuations, or statistical noise, in the
exponentially tiny number of the rare important configurations
completely washes out the significance of the result.

In layman's terms, imagine that we want to study ice, but can only run
experiments at normal room temperature and pressure. Using the
reweighting method is analogous to trying to glimpse the information
by waiting for rare configurations when all the water molecules
accidentally gather in one corner of the lab, forming a chunk of
ice. The amount of time that this experiment would require is
exponentially large as~$V\to\infty$.

\subsection{Theoretical predictions}
\label{sec:theory}

The first lattice prediction for the location of the critical point
has been reported in Ref.~\citen{FK}. The assumption is that, although
the problem becomes exponentially difficult as $V\to\infty$,
in practice, once can get a sensible approximation at finite $V$. 
In addition, simulations at finite $T$
might suffer lesser overlap problem because of large thermal 
fluctuations \cite{AKW}.
One can hope that if the critical point is at a small value of
$\mu_B$, the volume $V$ may not need to be too large
to achieve a reasonable accuracy. In particular, numerical estimates show
\cite{overlap} that the maximal value of $\mu_B$ which
one can reach within the same accuracy shrinks only as a power of
$1/V$.

However,
it is not possible to determine this accuracy, since the
exact result is unknown. Normally, one would estimate the error
by going to increasingly large volumes $V$, but, as discussed above,
the method becomes prohibitive too quickly (exponentially) in this
limit. Ultimately, the result of Ref.~\citen{FK} might turn out to be
a good approximation to the exact answer, but we can only tell
once we have an independent result to compare it to.
A qualitatively new approach is needed to overcome the QCD sign problem.
\footnote{In theories similar, or approximating, the finite density
QCD, the sign and/or overlap problems have been tackled recently,
using various new methods see, e.g.,
Refs.~\citen{Chandrasekharan,Wiese,Ambjorn}.}

\begin{figure}[htbp]
  \begin{center}
    \includegraphics[width=.8\textwidth]{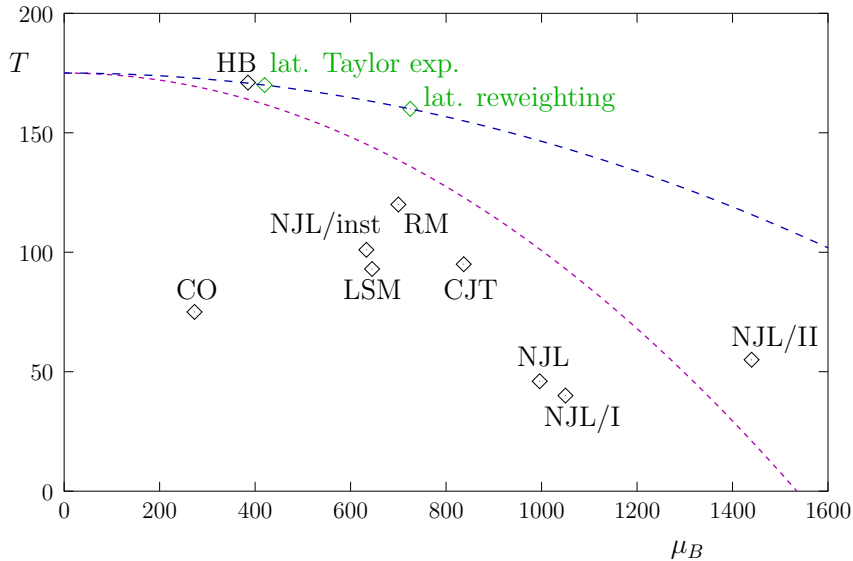}
    \caption[]{Theoretical  (models and lattice) 
predictions for the location of the critical point. The labels
correspond to Table \ref{tab:models}. The two dashed lines indicate
the magnitude of the slope $d^2T/d\mu^2$ obtained by lattice
Taylor expansion\cite{Ejiri}. The upper curve agrees with 
Ref.~\citen{Philipsen}.
The lower curve corresponds to smaller
quark mass. Errors/uncertainties are not shown.}
    \label{fig:tmudat}
  \end{center}
\end{figure}

In the absence of a controllable (i.e., systematically improvable)
method, one turns to model calculations. Many such calculations
have been done \cite{Asakawa,Barducci:1989,Barducci:1993,BergesRajagopal,Halasz,Scavenius,Antoniou,HattaIkeda}.  Figure \ref{fig:tmudat} 
and Table \ref{tab:models}
summarize the results. One can see that the predictions vary
wildly. An interesting point to keep in mind is 
that each of these models is tuned to
reproduce vacuum, $T=\mu_B=0$, phenomenology.
Nevertheless, extrapolation to nonzero $\mu_B$ 
is not constrained significantly by this. In a loose sense,
the existing lattice methods can be also viewed as extrapolations
from $\mu_B=0$, but finite $T$.

Two new lattice approaches are being developed. Each of them has
the capacity to determine the location of the critical point.
 One approach is based on
simulations at finite {\it imaginary} values of $\mu_B$
\cite{Philipsen} and the other on Taylor expansions around $\mu_B=0$
\cite{Ejiri}.
The curvature of the phase transition line found using these methods
is indicated by the upper parabola in Fig.\ref{fig:tmudat}.  Recent
 result \cite{Ejiri} (lower parabola in  Fig.\ref{fig:tmudat})
seems to indicate large sensitivity of this curvature 
to the quark mass. This may or may not be related to the
strong sensitivity of the position of the critical point to the
value of the strange quark mass observed in Ref.~\citen{Philipsen}.
Qualitatively, one should expect that the critical point
moves toward smaller $\mu_B$ as the strange quark mass $m_s$ is decreased,
since for sufficiently small $m_s$ the chiral transition must turn first order
according to renormalization group arguments\cite{PisarskiWilczek}.

\begin{table}[htbp]
  \begin{center}
\begin{tabular}{|l|r|l|l|}
\hline
Source &  $(T,\mu_B)$, MeV & Comments & Label\\
\hline\hline
MIT Bag/QGP & none & \em only 1st order, no chiral symmetry &---\\
\hline
Asakawa,Yazaki '89 &  (40, 1050) &  NJL, CASE I & NJL/I\\
\hline
`` &  (55, 1440) &  NJL, CASE II & NJL/II\\
\hline
Barducci, {\it et al} '89-94 & (75, 273)$_{\rm TCP}$ & composite operator &CO\\
\hline
Berges, Rajagopal '98 & (101, 633)$_{\rm TCP}$ & instanton NJL &NJL/inst\\
\hline
Halasz, {\it et al} '98 & (120, 700)$_{\rm TCP}$ & random matrix&RM\\
\hline
Scavenius, {\it et al} '01 & (93,645) & linear $\sigma$-model & LSM\\
\hline
`` & (46,996) & NJL & NJL\\
\hline
Fodor, Katz '01 & (160, 725) & lattice reweighting&\\
\hline
Hatta, Ikeda, '02 & (95, 837) & effective potential (CJT) & CJT\\
\hline
Antoniou, Kapoyannis '02 & (171, 385) & hadronic bootstrap & HB\\
\hline
Ejiri, {\it et al} '03 & (?,420) & lattice Taylor expansion&\\
\hline
\end{tabular}
    \caption{Theoretical predictions
of the location of the critical point. The predictions for
tricritical point are indicated as `TCP'. The last column gives the
corresponding label on~Fig.~\ref{fig:tmudat}. }
    \label{tab:models}
  \end{center}
\end{table}

\section{Scanning QCD phase diagram}

Even though the exact location of the critical point is not
known to us yet, the available theoretical estimates strongly
indicate that the point is within the region of the phase diagram
probed by the heavy-ion collision experiments. This raises the
possibility to discover this point in such experiments \cite{SRS}.

\begin{figure}
  \begin{center}
    \includegraphics[width=\textwidth]{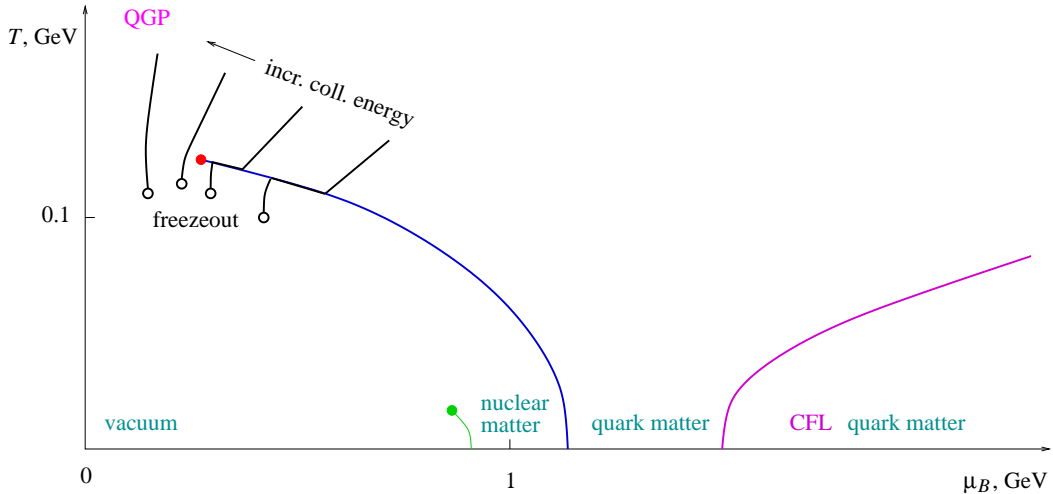}
    \caption[]{Example trajectories traced by a fireball created in a
    heavy ion collision on the phase diagram of Fig.\ref{fig:pdqcd}.
    Increasing the collision energy one moves the freezeout point
    (empty circle) to
    smaller $\mu_B$, approaching and then receding away from 
    the critical point.}
    \label{fig:pdqcd-fo}
  \end{center}
\end{figure}

The idea is illustrated in Fig.~\ref{fig:pdqcd-fo}. It is known empirically
that with increasing collision energy, $\sqrt s$, the resulting
fireballs tend to freezeout at decreasing values of the
chemical potential. This is easy to understand, since the
amount of generated entropy (heat) grows with $\sqrt s$
while the net baryon number is limited by that number in the initial nuclei.

The trajectories on Fig.~\ref{fig:pdqcd-fo} terminating in the
freezeout points indicate (theoretically perceived) time history of a
small, but thermodynamically macroscopic, volume of the expanding
fireball, at various initial collision energies. In the approximation
of ideal hydrodynamics these trajectories follow lines of constant
baryon per entropy ratio (baryon asymmetry), due to conservation of the
baryon number and the entropy.  The characteristic discontinuity of
the trajectory at the first order phase transition is a result of the
discontinuity of the baryon asymmetry across this transition. Because
the shift is toward the end point $E$, this leads to the phenomenon
of {\em focusing} \cite{SRS}:  the freezeout points tend to cluster near the
critical point for a wide range of initial trajectory
points. Therefore, wider range of $\sqrt s$ leads to freezeout in the
critical region, making the task of finding the point somewhat easier.

\begin{figure}
  \begin{center}
       \includegraphics[width=\halftext]{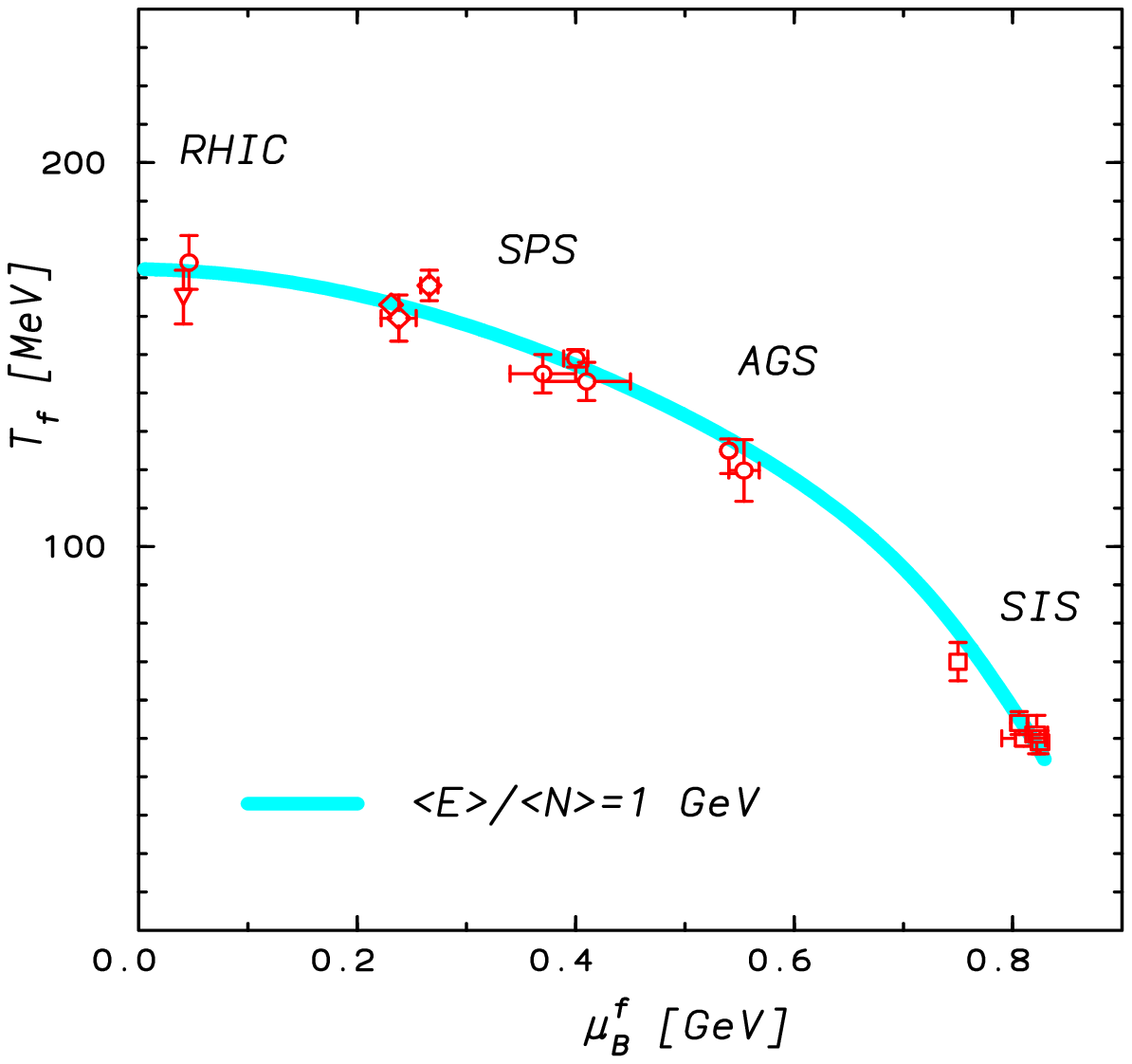}
       \includegraphics[width=\halftext]{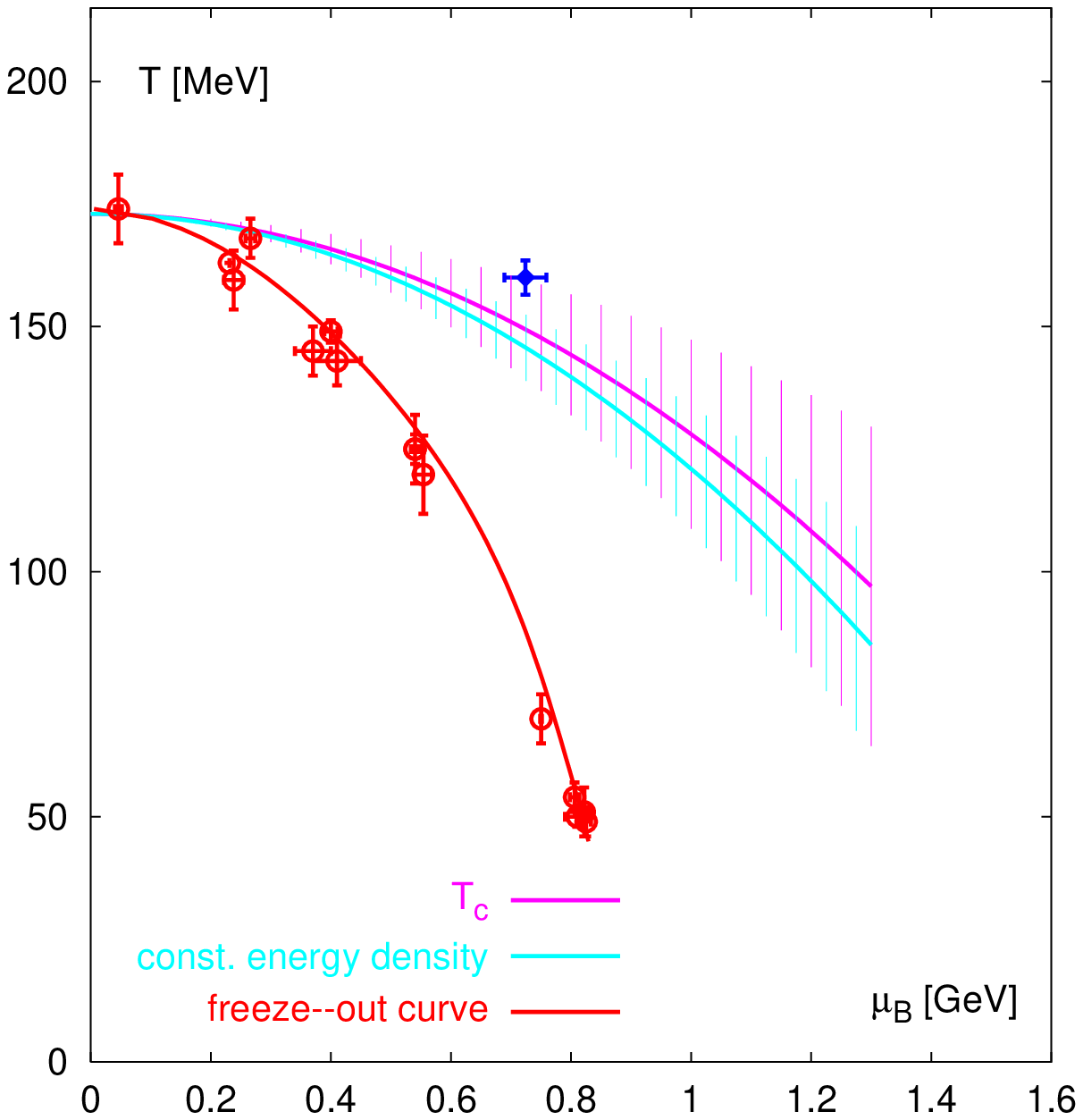}
    \caption[]{The freezeout points for different heavy-ion 
    collision experiments. On the right, the lattice results
\cite{FK,Ejiri,Philipsen} are superimposed.
Figures are reproduced from Ref.~\citen{BMRS}.}
    \label{fig:freezeout}
  \end{center}
\end{figure}

The information about the location of the freezeout point for given
experimental conditions
 is obtained by measuring the ratios of particle yields
(e.g., baryons or antibaryons to pions), and fitting to
a statistical model with $T$ and $\mu_B$ as parameters.
Such fits are amazingly good \cite{BMRS}, and
the resulting points for different experiments are shown in
Fig. \ref{fig:freezeout}.

For comparison, the location of the lattice reweighting calculation
result \cite{FK} is superimposed on the experimental freezeout curve
in Figure \ref{fig:freezeout}(right). It may appear that the critical
point is somewhat away (higher $T$) from the freezeout curve. However,
as emphasized already in Section \ref{sec:theory}, the {\em
systematic} error of the lattice result is not known, since the volume
$V\to\infty$ limit is unattainable using the reweighting method. Even
if one takes the lattice result at face value, one still has to take
into account the fact that the position of the critical point must
shift to smaller values of $\mu_B$ once the quark masses (notably, the
strange quark mass) are reduced toward their true values, from those
used in Ref.~\citen{FK}.  More recent lattice results \cite{Ejiri,Philipsen}
support this expectation.  The results of the model calculations (see
Figure \ref{fig:tmudat} and Table \ref{tab:models}) are closer to the
freezeout curve than the band in Fig.  \ref{fig:freezeout}.

Additional effect, which plays a significant role, is the critical
slowing down near the point $E$ \cite{SRS2,BR,DynScaling}. 
This phenomenon is the
major limiting factor (size limitation is less stringent
\cite{SRS2}) for the maximal correlation length that can be achieved
realistically in a heavy ion collision experiment. Although, as a
result, the sharpness of the signatures of the critical point is
reduced, another consequence is the shift of the position (due to
delay) of the maximum of the correlation length toward {\em lower}
temperatures \cite{BR}.

\begin{wrapfigure}{r}{\halftext}
  \begin{center}
\includegraphics[width=\halftext]{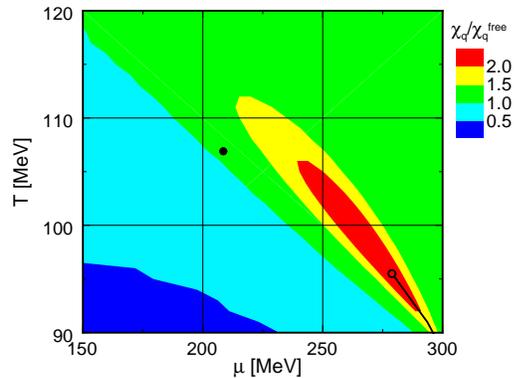}
    \caption[]{The model calculation \cite{HattaIkeda} of the shape of the
critical region. Note that $\mu=\mu_B/3$.}
    \label{fig:ikeda}
  \end{center}
\end{wrapfigure}

On the experimental side, one has some control over the freezeout
temperature by adjusting the size of the ions. Smaller systems
freeze out somewhat earlier (higher $T$). Also,
due to the singularity in the specific heat, the freezeout
occurs at higher $T$ for trajectories passing near the critical point
 \cite{SRS}.

Another interesting observation, potentially important for the search of the
critical point is that of the shape of the critical region \cite{HattaIkeda}.
One can see in Figure \ref{fig:ikeda} a model calculation of a
divergent susceptibility near the critical point, which shows that
the critical region is stretched in the direction of the 
crossover line.
This shape is easy to understand remembering that, in the $(T, \mu_B, m_q)$
space (see Fig.~\ref{fig:wings}), the critical point is 
connected to the tricritical point
(the black dot in Fig.~\ref{fig:ikeda}) by a whole line of critical points.

\section{Signatures: event-by-event fluctuations}

\newcommand\Free{{\cal F}}

One of the actively pursued signatures of the critical point is the
non-monotonous dependence on $\sqrt s$ (and thus, on $\mu_B$) of the
event-by-event fluctuation observables \cite{SRS,SRS2}.  The idea can
be understood qualitatively by noting that: (1) the
susceptibilities diverge at the critical point, and (2) the magnitude
of the fluctuations are proportional to the corresponding
susceptibilities. For example, for the fluctuations of energy or
charge, the well-known relations are
\begin{equation}
\frac{\partial E}{\partial T}
=\frac1{T^2}\langle(\Delta E)^2\rangle;\qquad
\frac{\partial Q}{\partial\phi}
= \frac1{T}\langle (\Delta Q)^2\rangle.
\end{equation}

Ideally, one could determine susceptibilities on the left-hand sides 
by measuring the fluctuations on the right-hand side 
\cite{C_V}. However,
practically, the measurement of the corresponding fluctuations, $\Delta
E$ or $\Delta Q$, is not feasible because not all the particles end up
in the detector \cite{SRS2,subevent}.  A more differential measure of the
fluctuations needs to be computed in theory and compared to
experiment.

\subsection{Two-particle correlator}

A number of such measures can be obtained starting from a two
particle correlator
\begin{equation}\label{dndn}
  \langle \Delta n^\alpha_{\bm p} \Delta n^\beta_{\bm k}\rangle
=
 \langle  n^\alpha_{\bm p}  n^\beta_{\bm k}\rangle
- 
 \langle  n^\alpha_{\bm p}\rangle   \langle n^\beta_{\bm k}\rangle
\end{equation}
where $\Delta n^\alpha_{\bm p} 
= n^\alpha_{\bm p} - \langle n^\alpha_{\bm p}\rangle$ 
is the event-by-event fluctuation of the number of particles of the
type $\alpha$ in the momentum bin centered around $\bm p$. 
 Experts familiar with Hanbury-Brown-Twiss (HBT) interferometry
\cite{HBT-review} may recognize in (\ref{dndn}) the HBT correlation
function. 

The two-particle correlator (\ref{dndn}) can be directly measured.
However, for such a function of many variables, it might
be difficult to represent the result of this measurement.
A useful representation, for example, is obtained by limiting
(projecting) the
variables to transverse components of $\bm p$ and $\bm k$.
The resulting plot of a function of two arguments, $p_T$ and $k_T$, is 
often referred to as a `Trainor plot' (see, e.g.,
 Ref.~\citen{NA49}). Interesting information can be also obtained by
 projecting onto the rapidities of $\bm p$ and $\bm k$. If in addition,
one weights each particle with its charge, the resulting
 correlator, as a function of the rapidity difference 
$y_{\bm p}-y_{\bm k}$, is essentially the balance function introduced
 in \cite{balance}.

There also exist many {\em cumulative} measures, proposed by theorists
and/or used by experimentalists,
\cite{Phi_pt,Mrow,SRS2,JeonKoch-ratio,Voloshin-lbnl,subevent,moments,charge,Jeon-review}
that can be expressed in terms of correlator (\ref{dndn}). As an
example, the fluctuation of electric charge is given by summing over
momenta $\bm p$ and $\bm k$ of {\em all} particles in the experimental
acceptance window and weighting each particle with its charge
$q^\alpha$:
\begin{equation}\label{deltaQ}
 {\Delta Q=\sum_{\bm p,\alpha} q^\alpha \Delta n_{\bm p}^\alpha} ;
\quad \mbox{thus} \qquad
\langle(\Delta Q)^2\rangle
=\sum_{\bm p,\alpha} \sum_{\bm k,\beta} 
q^\alpha q^\beta 
\langle \Delta n_{\bm p}^\alpha\Delta n_{\bm k}^\beta\rangle.
\end{equation}
The same applies to the fluctuations of the baryon charge, with
$q^\alpha$ substituted by the baryon charge of the particles. Similar
equation  (see Eq.(\ref{sigmaebe}))
also applies to the fluctuations of the mean 
transverse momentum $p_T$, in which case $q^\alpha$ should be replaced
with $p_T-\overline{p_T}$ -- the deviation of the momentum $p_T$ from
the all-event (inclusive) mean $\overline{p_T}$.

The correlator (\ref{dndn}) can,
in principle, be calculated, under assumption of thermal equilibrium,
once the relevant interactions are known.
In the case of the critical point, we need to concern ourselves
with the interactions which can lead to singular contribution
to the correlator (and, as a consequence, to susceptibilities)
as the critical point is approached.

In a non-interacting gas in thermal equilibrium the correlator
(\ref{dndn}) vanishes unless $\bm p=\bm k$ and $\alpha=\beta$.%
\footnote{We are 
not considering HBT correlations, which are a finite size effect.}
The hadrons, however, are interacting. One can ask a question: what is the
effect of the interaction on the correlator (\ref{dndn})?  The answer can be
found to leading order \cite{corr}. The contribution is proportional to
the amplitude of the forward scattering ${\cal A}_{pk\to pk}$ 
of the particles with momenta
$\bm p$ and $\bm k$. This is easy to understand using the following argument.
The amplitude of the forward scattering shifts the energy of the
2-particle state relative to the sum of single particle energies.
The statistical weight of the two particle state is therefore
changed relative to the product of the single-particle weights.
The difference is the two-particle correlator:

\begin{equation}
  \langle n_pn_k\rangle-\langle n_p\rangle\langle n_k\rangle 
= f_p f_k (e^{-\beta E_I} - 1)
\approx f_p f_k (-\beta E_I) \sim f_p f_k \beta\,{\cal A}_{pk\to pk}.
\end{equation}
where $f_p$ is the equilibrium distribution function and $E_I$ is the
interaction energy.
The exact formula, obtained using diagrammatic analysis,
\cite{corr} contains additional factors $(1+f_{\bm p})(1+f_{\bm k})$, which
can be understood as Bose enhancement (stimulated emission) factors
(or, in the case of fermions, $(1-f_{\bm p})(1-f_{\bm k})$ -- Pauli blocking).

Near the critical point the most singular contribution comes from the 
exchange of the sigma field quanta in the $t$ channel.%
\footnote{Strictly speaking, what we call here, for simplicity, 
``sigma'' is a mixture
(a linear combination) of chiral condensate, baryon density and energy density
fluctuations.}
 Since, by kinematics, the quanta carry zero momentum, the singular
contribution is proportional to $1/m_\sigma^2$, which equals $\xi^2$ --
the square of the sigma field correlation length.

\begin{wrapfigure}{l}{\halftext}
  \begin{center}
    \includegraphics[width=.25\textwidth]{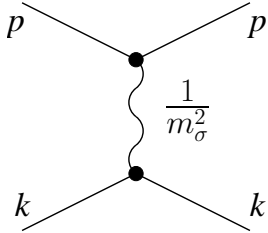}
    \caption[]{Diagrammatic representation of the singular
      contribution to the correlator 
      $\langle\Delta n_{\bm p} \Delta n_{\bm k}\rangle$.}
    \label{fig:t-sigma}
  \end{center}
\end{wrapfigure}

The absolute strength of the singularity depends on the coupling of
the critical mode sigma to the corresponding hadron in
Fig. \ref{fig:t-sigma}, which is
difficult to estimate reliably. Order of magnitude estimates
have been made for coupling to pions \cite{SRS2} and to protons
\cite{HattaS}\,.

As an example of the singular contribution in Fig.~\ref{fig:t-sigma}
consider baryon number susceptibility. Let $Q$ in 
equation (\ref{deltaQ}) be the baryon number. Then
one can see that the $1/m_\sigma^2$, or $\xi^2$, singularity
from Fig.~\ref{fig:t-sigma} for scattering two baryons results in the
divergence of the baryon number susceptibility (\ref{Bsusc}) ($\eta=0$
at this order).
If only charged baryons are detected,
the total baryon number cannot be measured event by event, 
but the number of protons is measurable. Since,
according to Fig. \ref{fig:t-sigma}, the proton number fluctuations
should also be singular at the critical point, measurement of such
fluctuations may provide a signal of the critical point \cite{HattaS}.

In principle, knowing the correlator (\ref{dndn}) one could make
 quantitative predictions for fluctuation measures used in
 experiment. In practice, calculating the correlator is a very
 difficult task (what interactions should be included and
 what is their strength?).
 Non-equilibrium effects make this task even more difficult.  Near the
 critical point these complications become less relevant since, as
 long as we limit ourselves to the singular effects, we only need to
 consider contributions such as in Fig.~\ref{fig:t-sigma}.

\subsection{Fluctuations, correlations, and acceptance}
\def\<{\langle}
\def\>{\rangle}
\def\p{{\bm{p}}}
\def\k{{\bm{k}}}
\def\q{{\bm{q}}}
\def\O{{\cal O}}
\def\yacc{{y_{\rm acc}}}

Cumulative measures of fluctuations are often used to represent
experimental results.
These measures suffer an important drawback -- they depend on the
size and shape of the acceptance window of the detector. This makes
comparison of different experiments, as well as an experiment to a
theory, difficult. However,
knowing certain properties of the correlator (\ref{dndn}), it is
possible to correct for acceptance in such comparisons.

As an illustration consider event-by-event
fluctuations of the mean transverse momentum $p_T$ per particle. 
Most commonly used fluctuation measures
are based on the width of the distribution of 
the event mean $p_T$,
$\sigma^2_{\rm ebe}$. Similar to (\ref{deltaQ}), $\sigma^2_{\rm ebe}$ 
can be expressed through
the correlator (\ref{dndn}) \cite{SRS2}:
\begin{equation}\label{sigmaebe}
 \sigma^2_{\rm ebe}
= \frac1{\langle N\rangle^2} \sum_{\bm p\bm k} 
\,{\Delta p_T \Delta k_T}
\,\<\Delta n_\p\,\Delta n_\k\>,
\end{equation}
where $\Delta p_T \equiv p_T - \overline{p_T}$ and $\<N\>$ is the
average multiplicity of accepted particles. In the thermodynamic limit
$\<N\>\to\infty$
the fluctuation $\sigma^2_{\rm ebe}$ vanishes as $1/\<N\>$. Thus
in this limit,
the  quantity $\<N\>\sigma^2_{\rm ebe}$ does not depend on the size 
of the system $\<N\>$  and is therefore a natural subject of 
theoretical predictions.

To make closer comparison to experiment, it is useful to
exclude the diagonal terms $\bm p=\bm k$ from the sum in
(\ref{sigmaebe}), since they give the trivial statistical contribution 
$\<N\>^{-1}\sigma_{\rm inc}^2$, where $\sigma_{\rm inc}$ is the
r.m.s. width of the inclusive distribution of $p_T$. The remaining
off-diagonal terms in (\ref{sigmaebe}) give the nontrivial 
``dynamical fluctuation'', experimentally obtained after the subtraction:
\begin{equation}
  \sigma_{\rm dyn}^2\equiv\sigma^2_{\rm ebe}-\<N\>^{-1}\sigma_{\rm inc}^2.
\end{equation}

In an experiment, the sum in (\ref{sigmaebe}) is limited to
$\bm p$ and $\bm k$ which fall within detector acceptance.
Assume, for clarity, that the acceptance is limited in rapidity,
i.e., $y_{\bm p}$ and $y_{\bm k}$ belong to an interval 
$[y_{\rm min},y_{\rm max}]$. The cumulative measure $\sigma^2_{\rm
  ebe}$, or $\sigma_{\rm dyn}^2$, 
will then depend on $y_{\rm acc}\equiv y_{\rm max}-y_{\rm
  min}$. This dependence simplifies in two regimes of $y_{\rm acc}$.
The boundary between the two regimes is determined by the
characteristic range $y_{\rm corr}$ 
of the rapidity correlator of the fluctuations:
\begin{equation}
  \<N\>\sigma^2_{\rm dyn}\Big|_{\yacc} =
\left\{ \begin{array}{ll}
O(\yacc), & \mbox{ for } \yacc\ll y_{\rm corr};\\\\
 \<N\>\sigma^2_{\rm dyn}\Big|_{\infty},
& \mbox{ for } \yacc\gg y_{\rm corr}.
\end{array}
\right.
\end{equation}
In other words, cumulative measure $\<N\>\sigma^2_{\rm dyn}$
grows linearly with $\yacc$ for small acceptance windows
and saturates at its thermodynamic limit value when the acceptance
range exceeds the correlation range. 
In most current experiments, the width of the rapidity window 
$\yacc$ is smaller
or at most comparable to the typical range of the rapidity correlator
$y_{\rm corr}\sim 1$. This means that in a typical experiment, 
for a cumulative measure, normalized to be finite in the
thermodynamic limit,
the experimentally observed
magnitude is roughly proportional to the acceptance window size 
\cite{Voloshin-lbnl,JeonPratt,Pruneau}.

\subsection{Experimental results and concluding remarks}

\begin{wrapfigure}{l}{\halftext}
          \includegraphics[width=\halftext]{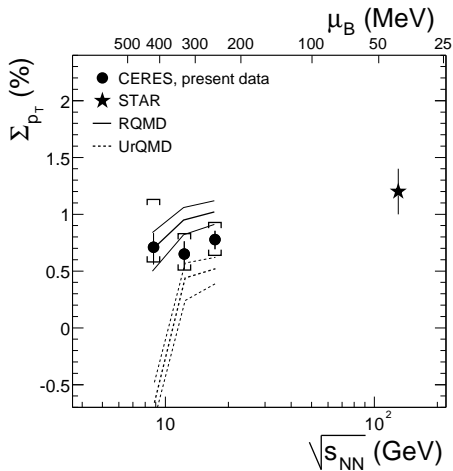}
      \caption[]{CERES and STAR results for different collision
	energies \cite{ceres}.
	The corresponding values of $\mu_B$ at freezeout are determined
      using statistical model analysis \cite{muB-vs-s}.}
      \label{fig:ceres}
\end{wrapfigure}

As an example of the QCD phase diagram scan, 
the plot in Fig.\ref{fig:ceres} shows the results of the
measurements of the $p_T$ fluctuations using a cumulative measure
$\Sigma_{p_T}$ described in Ref.~\citen{ceres}.
No clear
non-monotonous signal, which one would expect if the experiments
probed the vicinity of the critical point, was found.

It is also interesting
to compare the magnitude of the observed fluctuations to the singular
contribution expected
near the critical point \cite{SRS2}. After correcting for acceptance using the
method outlined in the previous section one finds:
\begin{equation}
\Sigma_{p_T}\sim 2\% \times { \left(\frac G{300\ {\rm MeV}}\right)^2
\left(\frac{\xi}{3 \ {\rm fm}}\right)^2}\,,
\end{equation}
where  $G$ is the magnitude of the $\sigma\pi\pi$ coupling in the
diagram Fig. \ref{fig:t-sigma} and $\xi=1/m_\sigma$.

It is important to note, that observation of a large magnitude of
fluctuations would not by itself constitute the signal of the critical
point. There are many possible contributions to the fluctuations,
which are difficult to estimate. The distinct signature of the critical
point is the non-monotonous behavior of fluctuation observables.

Experiments at other
energies, at CERN SPS, RHIC, and future GSI facility,
will be able to provide a complete
scan of the reachable domain on the QCD phase diagram and either
discover or rule out the presence of the critical point in this
domain. 

This review focused mainly on the signatures of the QCD critical point
based on the event-by-event fluctuations. Further study of the
properties of the critical point may reveal other, perhaps, even more
sensitive and experimentally cleaner signatures. For example,
real-time correlation functions and non-equilibrium dynamics 
near the critical point 
deserve further investigation \cite{Fukushima}.

Finally, the lack of a controllable and reliable theoretical method
to calculate coordinates of the critical point impairs our ability
to perform a more focused search. It is hard
to overemphasize the importance of such a theoretical
method.

\section*{Acknowledgements}

The author would like to thank the organizers of the {\em Finite
density QCD\/} workshop at Nara. He is grateful to Professor
T. Kunihiro, Professor A. Nakamura and to The Yukawa Institute for
Theoretical Physics for support and hospitality.

RIKEN BNL Center and U.S. Department of Energy [DE-AC02-98CH10886]
provided facilities essential for the completion of this work.
This work was supported in part by DOE grant No.\ DE-FG0201ER41195 
and by the Alfred P.\ Sloan Foundation.


%

\end{document}